\documentclass[prd,nofootinbib,twocolumn]{revtex4}
\usepackage{graphicx}
\usepackage{bm}
\usepackage{float}
\usepackage{subfigure}
\usepackage{amsmath}
\usepackage{amsfonts}
\usepackage{color}
\usepackage{slashed}
\usepackage{fancyhdr}

\def\be{\begin{equation}}
\def\ee{\end{equation}}
\def\bed{\begin{description}}
\def\eed{\end{description}}

\def\bea{\begin{eqnarray}}
\def\eea{\end{eqnarray}}

\def\ba{\begin{array}}
\def\ea{\end{array}}

\def\half{\frac{1}{2}\,}

\def\u1{$U(1)$}
\def\suu1{$SU(2)\times U(1)$}

\newcommand{\Ebar}{\bar{E}}

\newcommand{\Nc}{\mathcal{N}}

\begin{document}

\title{Fermions On One Or Fewer Kinks}

\author{Yi-Zen Chu$^1$ and Tanmay Vachaspati$^{1,2}$}
\affiliation{
$^1$CERCA, Department of Physics, Case Western Reserve University,
10900 Euclid Avenue, Cleveland, OH 44106-7079, \\
$^2$Institute for Advanced Study, Princeton, NJ 08540}

\begin{abstract}
\noindent We find the full spectrum of fermion bound states on a $Z_2$ kink. In
addition to the zero mode, there are ${\rm int}[2m_f/m_s]$ bound states, where
$m_f$ is the fermion and $m_s$ the scalar mass. We also study fermion modes on
the background of a well-separated kink-antikink pair. Using a variational
argument, we prove that there is at least one bound state in this background,
and that the energy of this bound state goes to zero with increasing
kink-antikink separation, $2L$, and faster than $e^{-a 2L}$ where $a={\rm
min}(m_s,2 m_f)$. By numerical evaluation, we find some of the low lying
bound states explicitly.
\end{abstract}

\maketitle

\section{Introduction}

A novel feature of fermion-topological defect interactions is the
appearance of fermion zero modes \cite{Caroli,Jackiw:1981ee,Vachaspatibook}.
The existence of zero modes has important implications, leading to
phenomena such as fractional quantum numbers \cite{Jackiw:1975fn} and
superconducting cosmic strings \cite{Witten:1984eb}. In any physical
setting, however, the system is expected to contain both defects and
antidefects, and extended topological defects will frequently occur as
closed structures, for example, closed loops of cosmic string, or
closed branes in brane cosmology. Then it is important to determine
the fate of a fermion zero mode in these situations.

The fate of fermion zero modes on topologically trivial structures,
such as kink-antikink or cosmic string loop, has been addressed
in Ref.~\cite{Postma:2007bf}. The expectation that the fermion
zero modes would be recovered as the kink-antikink separation,
or the size of the cosmic string loop, is increased indefinitely,
was not met in Ref.~\cite{Postma:2007bf}. In the present paper,
our primary aim is to reconsider the problem of fermions on
kink-antikink backgrounds. Contrary to Ref.~\cite{Postma:2007bf},
we find that there are bound states on kink-antikink pairs
whose energy vanishes exponentially fast with separation of the
kink and antikink.

We start by finding all fermion bound states on a single kink.
If $2m_f < m_s$ where $m_f$ and $m_s$ are the fermion and scalar
masses, we find that the bound state spectrum only contains a zero
mode. However, as we increase the fermion mass further, the number
of bound states increases and is bounded by $2m_f/m_s$ as
described in Sec.~\ref{boundstateonk}.
We then turn to the kink-antikink system, proving first
that a bound state exists if the kink and antikink are
well-separated. Our proof is based on a variational argument
and allows us to obtain an upper bound on the energy of
the bound state. The bound itself shows that the energy
goes to zero with separation ($2L$) faster than $\exp(-a2L)$
where $a={\rm min}(m_s,2m_f)$. Next, we evaluate the
bound state energies numerically and confirm the
exponential dependence on $L$. We also find an exponential
decay of the ground state energy with increasing $2m_f/m_s$.

In the next section we set up the problem. We summarize our
results in Sec.~\ref{summary}. Identities involving hypergeometric
function are included in the Appendix.

\section{Setup}
\label{setup}

The 1+1 dimensional field theory we are interested in is
described by the Lagrangian
\begin{equation}
L = \half (\partial_\mu \phi)^2 - \frac{\lambda}{4} (\phi^2-\eta^2)^2
    + i {\bar \psi} \gamma^\mu \partial_\mu \psi
    - g \phi {\bar \psi}\psi
\label{lagrangian}
\end{equation}
where $\phi$ is a real scalar field, $\psi$ is a two-component
spinor, and the $\gamma^\mu$ are defined as
\begin{equation}
\gamma^t = \sigma^3 = \left[ \begin{array}{cc}
                                 1&0 \\
                                 0&-1
                             \end{array} \right] \ , \ \
\gamma^z = i\sigma^1 = i \left[ \begin{array}{cc}
                                 0&1 \\
                                 1&0
                             \end{array} \right] \ , \ \
\end{equation}
There are two masses in the model. The scalar mass is
$m_s = \sqrt{2\lambda}\eta$ and the fermion mass is
$m_f = g\eta$, where we are taking $g >0$.

The $Z_2$ kink solution has the well-known form ({\it e.g.}
see Ref.~\cite{Vachaspatibook})
\begin{equation}
\phi = \eta \tanh\left(\frac{m_s z}{2} \right)
\label{kinkphisolution}
\end{equation}
and the antikink is obtained simply by letting $z\to -z$.
We shall also be interested in the system that contains
a well-separated kink and antikink, for which the scalar field
configuration can be chosen to be
\begin{equation}
\phi = \eta \tanh\left(\frac{m_s}{2}(z+L) \right)
      - \eta \tanh\left(\frac{m_s}{2}(z-L) \right) - \eta
\label{kinkantikink}
\end{equation}
The kink-antikink separation is $2L$.

Fermionic modes are found in the fixed scalar field background
by solving the Dirac equation,
\begin{equation}
\left( i\gamma^\mu\partial_\mu - g \phi \right) \psi = 0,
\label{dirac}
\end{equation}
where we will consider $\phi$ to be the kink solution of
Eq.~(\ref{kinkphisolution}) and the kink-antikink configuration in
Eq.~(\ref{kinkantikink}). The modes will contain a set of bound states ($|E| <
m_f$) and continuum states. In this paper, we will only be interested in
determining the bound states with $E>0$.

We write
\begin{equation}
\psi = e^{-iEt} \left[ \begin{array}{c}
(\beta_+ - \beta_-)/\sqrt{2}      \\
(\beta_+ + \beta_-)/\sqrt{2}      \\
\end{array} \right]
\end{equation}
to get
\begin{eqnarray}
(\partial_z + g\phi ) \beta_+ &=& - E \beta_-
\label{betap}\\
(\partial_z - g\phi ) \beta_- &=& + E \beta_+
\label{betaequations}
\end{eqnarray}

Before proceeding further, it is convenient to perform a change
to dimensionless variables defined by
\begin{eqnarray}
z' &=& \frac{m_s z}{2} \ , \quad
L' = \frac{m_s L}{2} \ , \quad
E' = \frac{2E}{m_s} \ ,
\nonumber \\
g' &=& \sqrt{\frac{2}{\lambda}} g  = \frac{2m_f}{m_s} \nonumber
\end{eqnarray}
In what follows, we will drop the primes for notational convenience.
The Dirac equations are then still given by Eqs.~(\ref{betap}),
(\ref{betaequations}),
though with all variables having their dimensionless meanings,
and the (rescaled) kink and kink-antikink backgrounds read
\begin{equation}
\phi_{\rm K} \equiv \tanh z
\end{equation}
\begin{equation}
\phi_{\rm K\overline{K}} \equiv \tanh (z+L)  - \tanh(z-L) - 1
\end{equation}

By substitution of one of Eqs.~(\ref{betap}), (\ref{betaequations})
into the other, we obtain the 1-dimensional Schr\"{o}dinger equations
for $\beta_\pm$,
\begin{equation}
-\partial_z^2 \beta_\pm +
        g (g \phi^2 \mp  \partial_z \phi ) \beta_\pm
               = E^2 \beta_\pm ,
\label{schrodinger}
\end{equation}
allowing us to identify the potentials
\begin{equation}
V_\pm(\phi) \equiv g(g \phi^2 \mp \partial_z \phi)
\end{equation}
Note that Eq.~(\ref{schrodinger}) actually contains two
Schr\"{o}dinger equations and the solutions of both must
yield the same eigenvalue $E^2$.

The single kink (and antikink)
backgrounds are odd functions of $z$, we see that under $z \to -z$,
their first order equations transform into
\begin{equation}
-\left( \partial_z \pm g \phi \right) \beta_{\pm}  = \mp E \beta_{\mp}.
\end{equation}
That is, the parity reserved positive energy solutions are the parity
un-reversed negative energy solutions. In other words, since kink and
antikink are parity reversed functions of each other, the positive energy
solutions on the kink are the negative energy solutions on the antikink;
the negative energy solutions on the kink are the positive energy solutions
on the antikink.
Further, since the derivative of an odd function is an even function
we observe that the corresponding Schr\"{o}dinger equation,
Eq.~(\ref{schrodinger}),  is invariant
under parity transformation: hence, if the energy eigenstates
turn out to be non-degenerate (they are, as we will see below),
they must be of a definite parity.

For even $\phi$, the first order equations (\ref{betap}),
(\ref{betaequations}) transform
under parity $z\to-z$ into
\begin{equation}
\left( \partial_z \mp g \phi \right) \beta_{\pm} = \pm E \beta_{\mp}.
\label{kkbarbetaeq}
\end{equation}
and hence $\beta_+(z) = \beta_-(-z)$. This includes the case of the
kink-antikink background. An alternate way to see this is that
$\partial_z \phi$ is an odd function of $z$, and the Schr\"{o}dinger
equation for $\beta_- (z)$ is identical to that for $\beta_+ (-z)$.
Hence if we have a solution to Eq.~(\ref{schrodinger}) for
$\beta_+ (z)$ for the kink-antikink background,
$\beta_- (z) = \beta_+ (-z)$ will be a solution for
the $\beta_-$ Schr\"{o}dinger equation with the same value of
$E^2$. In what follows, for the kink-antikink background, we will
simply work with the $\beta_+$ equation.

\section{Fermion bound states on a kink}
\label{boundstateonk}

We begin by solving the Schr\"{o}dinger equation for a fermion on a
single kink.
\begin{equation}
-\partial^2_z \beta_\pm + V_{K,\pm} (z) \beta_\pm = E^2 \beta_\pm
\label{singlekinkschrodinger}
\end{equation}
where
\begin{equation}
V_{K,\pm} (z) \equiv g^2 - g(g \pm 1) {\rm sech}^2 z
\label{kpot}
\end{equation}
For any value of $g > 0$, $V_{K,+}$ has the shape of a potential well with
asymptotic maximum of $g^2$, and minimum value of $-g$ at $z=0$. We know from
quantum mechanics in 1 dimension that every non-positive potential that
tends to zero asymptotically necessarily has at least one bound state. Hence
$V_{K,+} (z)$ has at least one bound state for every $g$. Also, since $V_{K,+}
(z)$ gets deeper with increasing $g$, we expect more and more bound states to
appear with larger values of $g$. This expectation will be confirmed below.
However, we also need a non-trivial bound state of the $\beta_-$
Schr\"{o}dinger equation which has the same energy eigenvalue as for $\beta_+$.
Only then will $\beta_\pm$ solve the first order equations,
Eq.~(\ref{betaequations}), except if $E=0$ for then we can take $\beta_-=0$.
For $0 < g \le 1$, $V_{K,-}$ is in the shape of a potential barrier and clearly
has no bound states. This shows that for $0< g\le 1$, the only possible bound
state is with $E=0$ and $\beta_-=0$; the solution is
\begin{equation}
\beta_+^{(0)} = {\rm sech}^g z
\end{equation}
More bound states do appear for $g > 1$ as we now find by explicit calculation.

Employing the prescription in Refs.~\cite{Vachaspatibook,morsefeshbach} we
write
\begin{equation}
\beta_\pm = \Nc_\pm {\rm sech}^bz F_\pm(z)
\end{equation}
with $b^2 = g^2 - E^2$, or $b = +\sqrt{g^2 - E^2}$, the positive
choice of sign to ensure square integrability. Next we
switch variables to
\begin{equation}
u \equiv \half (1-\tanh z)
\end{equation}
and obtain the hypergeometric equation,
\begin{eqnarray}
u(u-1) F''_\pm(u) + (b+1) (2u -1)F'_\pm(u) &+& \nonumber \\
\left( b(b+1) - g(g \pm 1) \right) F_\pm(u) &=& \hskip -0.1 cm 0
\end{eqnarray}
It can be inferred that the arguments of the hypergeometric function
$F[\alpha_\pm,\beta_\pm;\gamma_\pm;u]$ must be
\begin{eqnarray}
\alpha_\pm &=& b+\half - \left ( g\pm \half \right ) \nonumber \\
\beta_\pm &=& b+\half + \left ( g\pm \half \right ) \nonumber \\
\gamma_\pm &=& b + 1
\end{eqnarray}
Observe that the $(g\pm1/2)$ actually comes from taking a square root, so it
ought to be contained within an absolute value sign, $|g\pm1/2|$; but including
$\alpha$ and $\beta$ without the absolute value sign already covers both cases
$g\pm1/2>0$ and $g\pm1/2<0$, since the hypergeometric function obeys the
symmetry $F[\alpha_\pm,\beta_\pm;\gamma;u] = F[\beta_\pm,\alpha_\pm;\gamma;u]$.

The general solutions for $\beta_\pm$ are therefore
\begin{eqnarray}
\beta_\pm (z) &=&
C_1 {\rm sech}^b\ z ~
F[ \alpha_\pm, \beta_\pm; \gamma_\pm;u]
\nonumber \\
&& \hskip -2 cm + C_2 e^{bz}
F[ \alpha_\pm-\gamma+1, \beta_\pm-\gamma+1;2-\gamma;u]
\end{eqnarray}
As $z \to +\infty$, $\tanh z \to +1$ and from Eq.~(\ref{taylorhypergeometric})
the hypergeometric function after the $e^{bz}$ term goes to $1$. As a result,
we see that the second $C_2$ term becomes unbounded because of the $e^{bz}$
factor. Hence we need to set $C_2=0$ for normalizability.

As $z \to -\infty$, we use the identity in Eq.~(\ref{hypergeometricUto1-U})
to inform us that,
\begin{eqnarray}
\lim_{z\to-\infty} \beta_+(z) &=& \Nc_+ \biggl ( e^{bz}
\frac{\Gamma[b+1]\Gamma[-b]}{\Gamma[g+1]\Gamma[-g]}
\nonumber \\
&+& e^{-bz} \frac{\Gamma[b+1]\Gamma[b]}{\Gamma[b+g+1]\Gamma[b-g]} \biggr )
\end{eqnarray}
\begin{eqnarray}
\lim_{z\to-\infty} \beta_-(z) &=& \Nc_- \biggl( e^{bz}
\frac{\Gamma[b+1]\Gamma[-b]}{\Gamma[g]\Gamma[1-g]}
\nonumber \\
&+& e^{-bz} \frac{\Gamma[b+1]\Gamma[b]}{\Gamma[b+g]\Gamma[b-g+1]} \biggr )
\end{eqnarray}
The $e^{-bz}$ term would be unbounded if its coefficient is finite.
Recalling that the gamma function has poles at the negative
integers and zero, we can then set the $e^{-bz}$ term to zero by
requiring that the argument of one of the gamma functions in the
denominator to be a negative integer or zero.
Since both $b+g$ and $b+g+1$ are strictly positive, we need
\begin{equation}
b_n^\pm - g + \frac{1}{2} \mp \frac{1}{2} = -n_\pm \in \mathbb{Z}^-
\label{bnpm}
\end{equation}
which implies
\begin{eqnarray}
E_{n_+} = \sqrt{ n_+ (2g - n_+) } \nonumber \\
E_{n_-} = \sqrt{(n_- + 1)(2g - (n_- + 1))} \nonumber
\end{eqnarray}
The solution for $\beta_\pm$ is valid only if their energy
eigenvalues coincide, we get the additional requirement
\begin{equation}
n_+-n_- = +1
\end{equation}
The range of $n_+$ is determined by noting that $b_n^+ = g-n_+$ from
Eq.~(\ref{bnpm}) and normalizability requires $b_n^+ > 0$. Therefore
\begin{equation}
0 \le n_+ < g
\end{equation}

We then need to determine the relationship between the normalization constants
$\mathcal{N}_\pm$ of these $\beta_+$ and $\beta_-$ solutions by plugging them
back into our first order equations (\ref{betaequations}). With some algebra
involving the hypergeometric function identities
(\ref{hypergeometricderivative}) and (\ref{raisingandlowering}), we can verify
that our solutions do satisfy the first order equation provided we have
\begin{equation}
\frac{\Nc^{(n)}_+}{\Nc^{(n)}_-} = - \frac{E_n}{n}
\end{equation}
where $n=n_+$ labels the $n^{\rm th}$ mode.

To summarize, on the kink background the positive energy fermionic bound
states are given by
\begin{eqnarray}
\beta^{(n)}_+(z) &=& - \Nc_n \ E_n {\rm sech}^{g-n}z ~ \nonumber \\
&& \hskip -2 cm F\left[-n, 2g-n+1; g-n+1; \frac{1}{2}\left( 1 - \tanh(z)
\right) \right]
\end{eqnarray}
\begin{eqnarray}
\beta^{(n)}_-(z) &=& \Nc_n \ n \ {\rm sech}^{g-n}z ~ \nonumber \\
&& \hskip -2 cm F\left[-n+1, 2g-n; g-n+1; \frac{1}{2}\left( 1 - \tanh(z)
\right) \right]
\end{eqnarray}
\begin{equation}
E_n = \sqrt{n(2g-n)}, \qquad 0 \leq n < g, \ n \in\mathbb{Z}^+ \nonumber
\label{kinksolution}
\end{equation}
where we highlight that, because $-n$ and $-n+1$ are negative integers
or zero, we see from (\ref{taylorhypergeometric}) the hypergeometric
functions are
really finite order polynomials in $u = (1-\tanh~z)/2$.
\begin{eqnarray}
F\left[ -n, 2g-n+1; g-n+1; u \right] &=& \nonumber \\
&& \hskip -3 cm
\sum_{m=0}^n \frac{(-n)_m (2g+1-n)_m}{m!(g-n+1)_m} u^m \nonumber \\
F\left[-n+1, 2g-n; g-n+1; u \right] &=& \nonumber \\
&& \hskip -3 cm
\sum_{m=0}^{n-1} \frac{(-n+1)_m (2g-n)_m}{m! (g-n+1)_m} u^m
\nonumber
\end{eqnarray}

As an example, we can recover the bound state found in
Ref.~\cite{Postma:2007bf} by setting $n=1$,
\begin{eqnarray}
\beta^{(1)}_+(z) &=& -\Nc \sqrt{2g-1} \ {\rm sech}^{g-1} z ~ \tanh z
\nonumber \\
\beta^{(1)}_-(z) &=& \Nc {\rm sech}^{g-1} z
\nonumber \\
E_1 &=& \sqrt{ 2g - 1 }
\end{eqnarray}
where $\Nc$ is a normalization factor.

\section{Bound states on kink-antikink}

As discussed below Eq.~(\ref{kkbarbetaeq}), at the end of Sec.~\ref{setup},
it is sufficient to find the solution for $\beta_+(z)$ in the kink-antikink
background and then set $\beta_-(z) = \beta_+(-z)$. So we will only focus
on finding $\beta_+$.

\begin{figure}
\scalebox{0.5}{\includegraphics{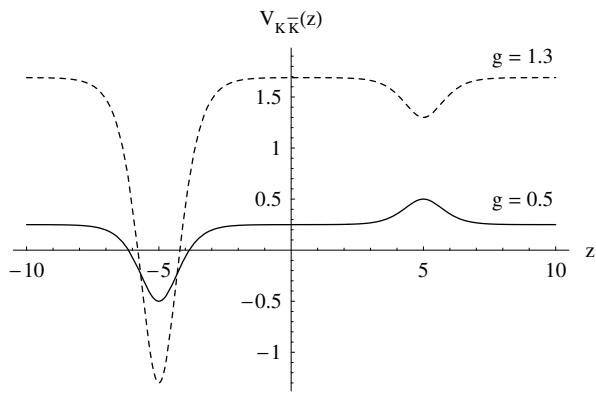}} \caption{Kink-antikink potentials
$V_{\rm K\overline{K}}$ for $g=0.5$ and $g=1.3$.} \label{VKKPlot}
\end{figure}

On inserting the kink-antikink background of Eq.~(\ref{kinkantikink}),
the Schr\"{o}dinger equation (\ref{schrodinger}) becomes
\begin{equation}
H_{K{\overline K}} \beta_+ \equiv
\left( -\partial^2_z +
V_{\rm K\overline{K}} \right) \beta_+ = E^2_n \beta_+
\label{kkbarschro}
\end{equation}
where the potentials are
\begin{eqnarray}
V_{\rm K\overline{K}} &\equiv&
V_{\rm K,+} + V_{\rm K,-} - g^2
\nonumber \\
&& + ~ 2 g^2 e^{-2L} {\rm sech}(z+L) ~ {\rm sech}(z-L) \label{kkbarpotential}
\end{eqnarray}
where the expressions for $V_{\rm K,\pm}$ are given in Eq.~(\ref{kpot}). The
shape of this potential is illustrated in Fig.~\ref{VKKPlot} for $g = 0.5$ and
$1.3$.

\subsection{Proof of existence of bound states}

There is a theorem by Simon \cite{simon} which states that a potential
$\epsilon V(z)$ admits at least one bound state for all $\epsilon > 0$ if and
only if $\int_{-\infty}^\infty V(z) dz \leq 0$.\footnote{An elementary proof by
computing the expectation value of the Hamiltonian with respect to some trial
wavefunction, can be found in \cite{brownstein}.} Applying this criterion to
our potentials (shifted by $-g^2$),
\begin{equation}
\int_{-\infty}^{+\infty} \left( V^{\rm (K\overline{K})}_\pm(z) -
g^2 \right) dz = -4 g^2 + 8 g^2 L \frac{e^{-2L}}{\sinh(2L)}
\end{equation}
At large $L$, $8 g^2 L e^{-2L}/\sinh(2L)$ is small compared to $4g^2$,
and hence the integral is negative. Solving for the zero of the
right hand side amounts to solving
\begin{equation}
4L + 1 = e^{4L}
\end{equation}
But $y = 4L + 1$ is the tangent line to $y = e^{4L}$ at $L=0$. That is,
the only solution to the above equation, and hence the only
instance the integral of the potential becomes non-negative, is when $L=0$.
For all $L > 0$, therefore, we see that the kink-antikink background,
as specified by Eq.~(\ref{kinkantikink}), supports at
least one fermion bound state for all non-zero values of the coupling $g$.
Contrary to the claim by Postma and Hartmann \cite{Postma:2007bf}, we see
that spin does not pose any obstacle to the existence of fermion bound states
on the kink-antikink.

\subsection{A lowest energy upper bound}

As mentioned in \cite{Postma:2007bf}, the fermion zero mode ($E = 0$) solution
on the kink-antikink is not normalizable, as can be verified by integrating
(\ref{betaequations}) directly. That means $E_0^2$ is strictly positive. From
the variational principle in quantum mechanics, we also know that the ground
state energy $E_0^2$ is always less than or equal to the expectation value of
the Hamiltonian $H_{K{\overline K}}$ with respect to an arbitrary square
integrable wavefunction $|\psi\rangle$, namely,
\begin{equation}
E_0^2 \leq \frac{\langle \psi | H_{K{\overline K}}
                     | \psi \rangle}{\langle \psi | \psi \rangle}
\label{varprinciple}
\end{equation}
Motivated by the fact that
\begin{equation}
\varphi(z) \equiv {\rm sech}^g(z+L)
\end{equation}
is the $\beta_+$ zero mode solution to a single kink at $z=-L$ and the only
normalizable $\beta_+$ solution to the antikink at $z=+L$ is
zero, we shall use $\varphi$ as our trial wavefunction.

Inserting the Hamiltonian in Eq.~(\ref{varprinciple}) and using the
equation obeyed by the zero mode state (Eq.~(\ref{singlekinkschrodinger})
with $E=0$) we get
\begin{eqnarray}
0 &<& E_0^2 \leq \frac{\Gamma\left[g+\frac{1}{2}\right]}{\sqrt{\pi}\Gamma[g]}
\int_{-\infty}^\infty dz ~ {\rm sech}^{2g}z_+ ~ {\rm sech}\ z_-
\nonumber \\
&\times &\biggl [ -g(g-1) {\rm sech}\ z_- + 2g^2 e^{-2L} {\rm sech}\ z_+ \biggr
] \label{E02InequalityA}
\end{eqnarray}
where we have denoted $z_\pm = z\pm L$ and also used the result
\cite{lebedev,hochstadt}
\begin{equation}
\int_{-\infty}^\infty {\rm sech}^{2g}z dz =
\frac{\sqrt{\pi}\Gamma[g]}{\Gamma\left[g+\frac{1}{2}\right]}
\end{equation}
The second term in the bracket in Eq.~(\ref{E02InequalityA})
gives a contribution proportional to
\begin{eqnarray}
2g^2 e^{-2L} \int dz && \hskip -0.2 cm {\rm sech}^{2g+1}z_+ ~ {\rm sech}\ z_-
\nonumber \\
&& < 8g^2 e^{-4L} \int dz~ e^z {\rm sech}^{2g+1} z
\end{eqnarray}
where we have used the inequality ${\rm sech}\ z_- < 2 e^{z_-}$. The first term
in the bracket also gives a contribution proportional to $e^{-4L}$ for $g > 1$.
However, for $0 < g < 1$, the contribution is estimated using
\begin{eqnarray}
g(1-g) \int dz && \hskip -0.2 cm {\rm sech}^{2g}z_+ ~{\rm sech}^2 z_-
\nonumber \\
&& \hskip -1 cm <  ~ g(1-g) 2^{2g} e^{-4gL} \int dz~ e^{2gz} {\rm sech}^2 z
\end{eqnarray}
where we have used the inequality ${\rm sech}^{2g} z_+ < 2^{2g} e^{2gz_+}$.

The end result is
\begin{equation}
0 < E_0^2 < e^{-4L}
\frac{\Gamma\left[g+\frac{1}{2}\right]}{\sqrt{\pi}\Gamma[g]}
8g^2 \int dz~ e^z {\rm sech}^{2g+1} z
\end{equation}
if $g > 1$, and
\begin{equation}
0 < E_0^2 < e^{-4gL}
\frac{\Gamma\left[g+\frac{1}{2}\right]}{\sqrt{\pi}\Gamma[g]}
g(1-g) 2^{2g} \int dz~ e^{2gz} {\rm sech}^2 z
\end{equation}
if $0 < g < 1$ in the large $L$ limit where the first term
in Eq.~(\ref{E02InequalityA}) dominates over the second term.

These results provide an upper bound for the energy of the
ground state in the kink-antikink background, the existence
of which we proved in the previous subsection.

\subsection{Numerical Solutions}

We proceed to numerically solve the fermion bound state on the kink-antikink.

First we note that it is impossible for $\beta_\pm$ to both vanish at the same
$z$. Recall that first order equations are solved uniquely by specifying one
boundary condition for each $\beta$. So if it were the case that $\beta_+(z_0)
= \beta_-(z_0) = 0$ for some $z_0$, then looking at (\ref{betaequations}), the
unique solution is simply $\beta_+(z) = \beta_-(z) = 0 \ \forall z$. In
particular, we cannot have both $\beta_\pm$ go to zero at $z=0$. As discussed
earlier, since $\beta_+(z) = \beta_-(-z)$ for the kink-antikink we can thus set
$\beta_\pm(z=0) = 1$ and rescale the solutions later if necessary.

The eigenvalues are written as $E_0 = \sqrt{|2g-1|}\delta$ and, for $n\geq
1$, $E_n = \Ebar_n (1+\delta)$, with $\Ebar_n \equiv \sqrt{n(2g-n)}$. They are
searched for by solving $(\ref{betaequations})$ repeatedly with various values
of $\delta$, and watching the large $|z|$ asymptotic behavior of the solutions,
as in the ``shooting method''. All of them eventually blow up, but as one tunes
$\delta$, the $\beta_+$ may say switch from going to negative infinity to going
to positive infinity, as $z \to -\infty$. The exact eigenvalue lies between
these two values of $\delta$ where this transition takes place, and the search
for the eigenvalue primarily involves narrowing the gap between these two
$\delta$s until the desired accuracy is achieved.

\begin{figure}
\scalebox{0.45}{\includegraphics{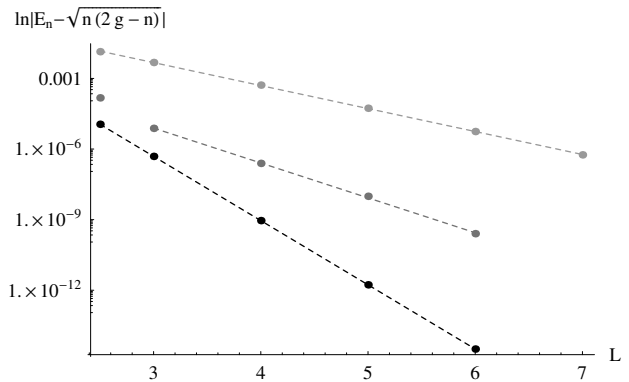}}
\caption{Ground state and excited energy levels of fermion on kink-antikink
near $\overline{E}_n = \sqrt{n(2g-n)}$, which are the energy levels on the
single kink, for $g=\pi$. Here we plot the absolute value of the deviation from
$\overline{E}_n$ to show, for the first three levels, the roughly exponential
dependence on $L$, {\it i.e.} $\delta E_n \equiv |E_n - \overline{E}_n| \sim
e^{-aL}$, with $a>0$. From dark to light, the dots are for $n=0,1$ and $2$,
with best-fit slopes of -6.28, -3.41 and -2.25 respectively.} \label{gPiPlot}
\end{figure}

We selected $g=\pi$ and investigated how the energy levels near those of the
single kink, $\sqrt{n(2\pi-n)}$, $n \in \{0,1,2\}$, are varied as the
kink-antikink separation is altered from $L=2.5$ thru $L=7$. Referring to
Fig.~\ref{gPiPlot}, one can infer that the first three energy levels roughly
have an exponential dependence on the kink-antikink distance: $E_n \sim
e^{-aL}$, for some $a>0$ dependent on $n$. This indicates the $\{E_n\}$
approach that of their single kink counterparts as $L$ is increased, in
accordance with physical intuition.

\begin{figure}
\scalebox{0.5}{\includegraphics{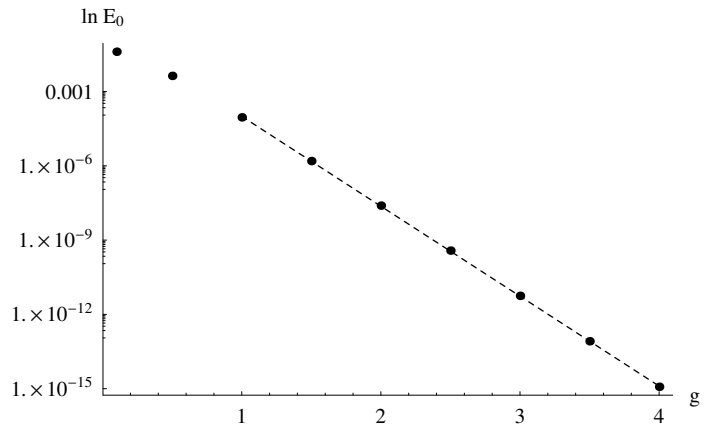}}
\caption{Ground state energy vs. $g$, the Yukawa coupling, for $L=5$. We see
that $E_0 \sim e^{-8.36 g}$.} \label{L5Plot}
\end{figure}

For $L=5$, we varied the coupling $g$ from $0.1$ thru $4$ to examine the effect
on the ground state energy eigenvalues. Fig.~\ref{L5Plot} provides evidence
that the energies decrease roughly exponentially with increasing strength of
the coupling.

\begin{figure}
\scalebox{0.6}{\includegraphics{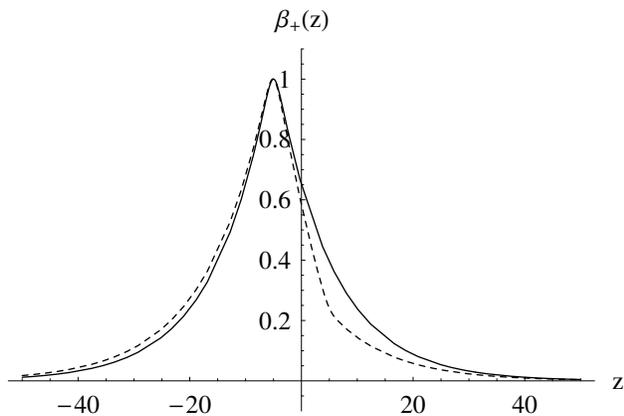}}
\caption{Ground state of fermion on kink-antikink with $g = 0.1$, $L = 5$, and
$E_0 \approx 0.04$. The solid line is the ground state $\beta_+ = {\rm
sech}^g(z+L)$ solution on a single kink centered at $z = -L$. The dashed line
is the numerical solution to the kink-antikink system.} \label{groundstatePlot}
\end{figure}

The remaining figure, Fig.~\ref{groundstatePlot}, shows the numerical
$\beta_+$ solution to the kink-antikink system for the ground state of
$\{g,\ L\} = \{0.1,\ 5\}$. It is compared against the corresponding
analytic solution $\beta_+(z) = {\rm sech}^g(z+L)$ for the single kink
at $z=-L$; the $\beta_+$ solution for the single antikink at $z=+L$
is zero. The numerical solution is normalized so that its approximate
peak at $z = -L$ coincides with that of ${\rm sech}^g(z+L)$.

\section{Conclusions}
\label{summary}

We have tackled the problem of solving for bound states of
the Dirac equation in (1+1) dimensions on kink and kink-antikink
backgrounds. The resulting coupled first order equations can in
turn be uncoupled to yield two Schr\"{o}dinger equations, which
we solve exactly for the single kink and antikink case. We find
that the number of positive energy bound states on a kink is given
by the smallest integer less than $g=2m_f/m_s$. For
fermions on a kink-antikink, we used the Schr\"{o}dinger equations
and results from non-relativistic
quantum mechanics to prove that at least one bound state has to exist,
for all non-zero values of the Yukawa coupling $g$. We then derived
an upper bound for the lowest energy squared $E_0^2$ value which allowed
us to prove that the ground state energy of the fermion on the
kink-antikink tends to zero as the kink-antikink separation tends
to infinity ($L \to \infty$). Appropriate boundary
conditions for the first order equations were devised and employed to
solve numerically the energy eigenvalues and eigenfunctions. For
the specific examples we looked at, the lower lying bound states
approached that of their single kink counterparts exponentially
quickly as the kink-antikink distance was increased. Similarly,
the ground state energy approached zero exponentially quickly as
one increased the strength of the Yukawa coupling.

We expect our results to be valid also for the case of vortex-antivortex pairs,
and for the case of loops of cosmic string. The lowest non-negative energy
state on a loop of cosmic string will have positive energy that is suppressed
by $\exp(-c R/w)$ where $R$ is the radius of the loop and $w$ is a width
associated with the string and $c$ is a numerical constant of order unity. In
cosmological applications, this is an enormous suppression and we expect the
picture derived on the assumption of exact zero modes to still hold true.
Exceptions could occur if a loop shrinks and becomes small, or where a cusp
occurs on a loop. For the case of superconducting strings \cite{Witten:1984eb},
the small but non-zero energy of the lowest positive energy state means that
charge carriers now have to jump from the Dirac sea to positive energy,
requiring $2m$ energy, where $m$ is the mass of the lowest positive energy
state. An applied electric field with strength $< m^2 /e$ along the string can
cause this jump as in Schwinger pair production but the process is due to
tunneling and is exponentially suppressed \cite{Schwinger:1951nm}. At stronger
electric fields, the process would be unsuppressed. The critical value of the
electric field for unsuppressed pair production is $\sim m_f^2 \exp(-cL/w) /e$
where $e$ is the electric charge of the fermion.

Another setting where fermion zero modes are believed to play an
important role is in brane cosmology where fermions are trapped
on 3+1 dimensional branes in a higher dimensional bulk universe.
If the fermions have zero modes in the brane background, it corresponds
to massless fermions that are trapped on the brane and this is a
possible explanation for massless standard model fermions living
in a 3 dimensional space. In light of our results, if the brane
can be thought of as a domain wall, in addition to the fermion zero
modes, we may also expect other bound states to exist for a range
of parameters. If the brane is closed or the bulk contains neighboring
antibranes, the fermion zero modes will become bound states with an
exponentially small mass. This may either be viewed as an undesirable
feature of the particular brane system, or else may be viewed as a
means to probe brane configurations in the bulk via the properties
of standard model fermions.

\begin{acknowledgments}
We thank Marieke Postma for discussions. Y.-Z.C. thanks Dai De-Chang for
discussing methods of solving (\ref{betaequations}) in the early stages of this
paper. All numerical and some analytic work in this paper were performed with
\emph{Mathematica} \cite{mathematica}. This work was supported in part by the
U.S. Department of Energy and NASA at Case Western Reserve University.
\end{acknowledgments}

\appendix

\section{Hypergeometric Function Identities}

In this appendix we collect various hypergeometric identities
\cite{AMS55,lebedev,morsefeshbach,temme} used in this paper.

\begin{eqnarray}
\label{taylorhypergeometric}
F[\alpha,\beta;\gamma;u] &=&
\sum_{m=0}^\infty \frac{(\alpha)_m (\beta)_m}{m!(\gamma)_m} u^m
\nonumber \\
&& \hskip -2.5 cm
(\sigma)_m \equiv (\sigma)(\sigma+1)\dots(\sigma+m-1), |u|<1
\end{eqnarray}
\begin{eqnarray}
\label{hypergeometricUto1-U}
F[\alpha,\beta;\gamma;u] &=& \nonumber \\
&& \hskip -2.5 cm
\frac{\Gamma[\gamma]\Gamma[\gamma-\alpha-\beta]}
      {\Gamma[\gamma-\alpha]\Gamma[\gamma-\beta]}
F[\alpha,\beta;1+\alpha+\beta-\gamma;1-u] \nonumber\\
&& \hskip -2 cm
+ (1-u)^{\gamma-\alpha-\beta}
\frac{\Gamma[\gamma]\Gamma[\alpha+\beta-\gamma]}{\Gamma[\alpha]\Gamma[\beta]}
\nonumber \\
&& \hskip -1 cm
\times F[\gamma-\alpha,\gamma-\beta;1-\alpha-\beta+\gamma;1-u], \nonumber \\
&&
|{\rm arg}[u]| < \pi,
\ |{\rm arg}[1-u]| < \pi, \nonumber \\
&& \alpha+\beta-\gamma \ \neq 0,\pm1,\pm2,\dots
\end{eqnarray}
\begin{equation}
\label{hypergeometricderivative} u \frac{d}{du}F[\alpha,\beta;\gamma;u] =
\alpha \left( F[\alpha+1,\beta;\gamma;u] - F[\alpha,\beta;\gamma;u] \right)
\end{equation}
\begin{eqnarray}
\label{raisingandlowering}
(\alpha+1 - \beta)(1-u)F[\alpha+1,\beta;\gamma;u] &=& \nonumber \\
&&
\hskip -5 cm
(\alpha+1 - \gamma) F[\alpha,\beta;\gamma;u] \nonumber \\
&&
\hskip -4 cm
+ (\gamma-\beta) F[\alpha+1,\beta-1;\gamma;u]
\end{eqnarray}

\end{document}